\documentclass[3p,times]{elsarticle}

\usepackage{ecrc}

\volume{00}

\firstpage{1}

\journalname{High Energy Density Physics}

\runauth{V. Dwarkadas et al.}

\jid{hedp}

\jnltitlelogo{High Energy Density Physics}

\CopyrightLine{2011}{Published by Elsevier Ltd.}

\usepackage{amssymb}

\usepackage[figuresright]{rotating}

\begin{document}
\newcommand{\vper}{\mbox{${v_{\perp}}$}}
\newcommand{\vpar}{\mbox{${v_{\parallel}}$}}
\newcommand{\uper}{\mbox{${u_{\perp}}$}}
\newcommand{\vperout}{\mbox{${{v_{\perp}}_{o}}$}}
\newcommand{\uperout}{\mbox{${{u_{\perp}}_{o}}$}}
\newcommand{\vperin}{\mbox{${{v_{\perp}}_{i}}$}}
\newcommand{\uperin}{\mbox{${{u_{\perp}}_{i}}$}}
\newcommand{\upar}{\mbox{${u_{\parallel}}$}}
\newcommand{\uparout}{\mbox{${{u_{\parallel}}_{o}}$}}
\newcommand{\vparout}{\mbox{${{v_{\parallel}}_{o}}$}}
\newcommand{\uparin}{\mbox{${{u_{\parallel}}_{i}}$}}
\newcommand{\vparin}{\mbox{${{v_{\parallel}}_{i}}$}}
\newcommand{\dout}{\mbox{${\rho}_{o}$}}
\newcommand{\din}{\mbox{${\rho}_{i}$}}
\newcommand{\da}{\mbox{${\rho}_{1}$}}
\newcommand{\mfast}{\mbox{$\dot{M}_{f}$}}
\newcommand{\mslow}{\mbox{$\dot{M}_{a}$}}
\newcommand{\beqn}{\begin{eqnarray}}
\newcommand{\eeqn}{\end{eqnarray}}
\newcommand{\be}{\begin{equation}}
\newcommand{\ee}{\end{equation}}
\newcommand{\noi}{\noindent}
\newcommand{\ftheta}{\mbox{$f(\theta)$}}
\newcommand{\gtheta}{\mbox{$g(\theta)$}}
\newcommand{\ltheta}{\mbox{$L(\theta)$}}
\newcommand{\stheta}{\mbox{$S(\theta)$}}
\newcommand{\utheta}{\mbox{$U(\theta)$}}
\newcommand{\xitheta}{\mbox{$\xi(\theta)$}}
\newcommand{\vs}{\mbox{${v_{s}}$}}
\newcommand{\ro}{\mbox{${R_{0}}$}}
\newcommand{\pa}{\mbox{${P_{1}}$}}
\newcommand{\va}{\mbox{${v_{a}}$}}
\newcommand{\vo}{\mbox{${v_{o}}$}}
\newcommand{\vp}{\mbox{${v_{p}}$}}
\newcommand{\vw}{\mbox{${v_{w}}$}}
\newcommand{\vf}{\mbox{${v_{f}}$}}
\newcommand{\lprime}{\mbox{${L^{\prime}}$}}
\newcommand{\uprime}{\mbox{${U^{\prime}}$}}
\newcommand{\sprime}{\mbox{${S^{\prime}}$}}
\newcommand{\xiprime}{\mbox{${{\xi}^{\prime}}$}}
\newcommand{\mdot}{\mbox{$\dot{M}$}}
\newcommand{\msun}{\mbox{$M_{\odot}$}}
\newcommand{\yr}{\mbox{${\rm yr}^{-1}$}}
\newcommand{\kms}{\mbox{${\rm km} \;{\rm s}^{-1}$}}
\newcommand{\lambdav}{\mbox{${\lambda}_{v}$}}
\newcommand{\lequ}{\mbox{${L_{eq}}$}}
\newcommand{\eqpratio}{\mbox{${R_{eq}/R_{p}}$}}
\newcommand{\ra}{\mbox{${r_{o}}$}}
\newcommand{\bfig}{\begin{figure}[h]}
\newcommand{\efig}{\end{figure}}
\newcommand{\tone}{\mbox{${t_{1}}$}}
\newcommand{\done}{\mbox{${{\rho}_{1}}$}}
\newcommand{\dsn}{\mbox{${\rho}_{SN}$}}
\newcommand{\dzero}{\mbox{${\rho}_{0}$}}
\newcommand{\ve}{\mbox{${v}_{e}$}}
\newcommand{\vej}{\mbox{${v}_{ej}$}}
\newcommand{\Mch}{\mbox{${M}_{ch}$}}
\newcommand{\mej}{\mbox{${M}_{e}$}}
\newcommand{\Mst}{\mbox{${M}_{ST}$}}
\newcommand{\dam}{\mbox{${\rho}_{am}$}}
\newcommand{\Rst}{\mbox{${R}_{ST}$}}
\newcommand{\Vst}{\mbox{${V}_{ST}$}}
\newcommand{\Tst}{\mbox{${T}_{ST}$}}
\newcommand{\no}{\mbox{${n}_{0}$}}
\newcommand{\Efif}{\mbox{${E}_{51}$}}
\newcommand{\rsh}{\mbox{${R}_{sh}$}}
\newcommand{\msh}{\mbox{${M}_{sh}$}}
\newcommand{\vsh}{\mbox{${V}_{sh}$}}
\newcommand{\vrev}{\mbox{${v}_{rev}$}}
\newcommand{\rpr}{\mbox{${R}^{\prime}$}}
\newcommand{\mpr}{\mbox{${M}^{\prime}$}}
\newcommand{\vpr}{\mbox{${V}^{\prime}$}}
\newcommand{\tpr}{\mbox{${t}^{\prime}$}}
\newcommand{\cone}{\mbox{${c}_{1}$}}
\newcommand{\ctwo}{\mbox{${c}_{2}$}}
\newcommand{\cthree}{\mbox{${c}_{3}$}}
\newcommand{\cfour}{\mbox{${c}_{4}$}}
\newcommand{\Te}{\mbox{${T}_{e}$}}
\newcommand{\Ti}{\mbox{${T}_{i}$}}
\newcommand{\Ha}{\mbox{${H}_{\alpha}$}}
\newcommand{\Rprime}{\mbox{${R}^{\prime}$}}
\newcommand{\Vprime}{\mbox{${V}^{\prime}$}}
\newcommand{\Tprime}{\mbox{${T}^{\prime}$}}
\newcommand{\Mprime}{\mbox{${M}^{\prime}$}}
\newcommand{\rprime}{\mbox{${r}^{\prime}$}}
\newcommand{\rfprime}{\mbox{${r}_f^{\prime}$}}
\newcommand{\vprime}{\mbox{${v}^{\prime}$}}
\newcommand{\tprime}{\mbox{${t}^{\prime}$}}
\newcommand{\mprime}{\mbox{${m}^{\prime}$}}
\newcommand{\Me}{\mbox{${M}_{e}$}}
\newcommand{\nh}{\mbox{${n}_{H}$}}
\newcommand{\rr}{\mbox{${R}_{2}$}}
\newcommand{\rf}{\mbox{${R}_{1}$}}
\newcommand{\vtwo}{\mbox{${V}_{2}$}}
\newcommand{\vout}{\mbox{${V}_{1}$}}
\newcommand{\dshell}{\mbox{${{\rho}_{sh}}$}}
\newcommand{\dwind}{\mbox{${{\rho}_{w}}$}}
\newcommand{\dslow}{\mbox{${{\rho}_{s}}$}}
\newcommand{\dfast}{\mbox{${{\rho}_{f}}$}}
\newcommand{\vfast}{\mbox{${v}_{f}$}}
\newcommand{\vslow}{\mbox{${v}_{s}$}}
\newcommand{\cc}{\mbox{${\rm cm}^{-3}$}}
\newcommand{\apj}{\mbox{ApJ}}
\newcommand{\apjl}{\mbox{ApJL}}
\newcommand{\apjs}{\mbox{ApJS}}
\newcommand{\aj}{\mbox{AJ}}
\newcommand{\araa}{\mbox{ARAA}}
\newcommand{\nat}{\mbox{Nature}}
\newcommand{\aap}{\mbox{AA}}
\newcommand{\gca}{\mbox{GeCoA}}
\newcommand{\pasp}{\mbox{PASP}}
\newcommand{\mnras}{\mbox{MNRAS}}
\newcommand{\apss}{\mbox{ApSS}}

\begin{frontmatter}



\dochead{}

\title{On the Evolution of the Maximum Energy of Accelerated Particles
  in Young Supernova Remnants}


\author[vvd]{V. V. Dwarkadas\corref{cor1}}
\ead{vikram@oddjob.uchicago.edu}

\author[it]{I. Telezhinsky}
\ead{telezhinsky@gmail.com}

\author[mp]{M. Pohl}
\ead{pohlmadq@gmail.com}

\address[vvd]{Department of Astronomy and astrophysics, Univ of Chicago, TAAC 55, Chicago, IL 60637}

\address[it]{Deutsches Elektronen-Synchrotron, Platanenallee 6, 15738
  Zeuthen, Germany}

\address[mp]{Institut fur Physik \& Astronomie, Universitat Potsdam,
  14476 Potsdam, Germany}

\begin{abstract}

It has generally been assumed in the literature that while young
supernova remnants (SNRs) accelerate particles even in the early
stages, the particles do not escape until the start of the
Sedov-Taylor or adiabatic stage, when the maximum energy of
accelerated particles is reached. These calculations however do not
take into account the detailed hydrodynamical expansion in the
ejecta-dominated stage, and the approach to the Sedov stage. Using
analytic approximations, we explore different environments in which
the SNR may evolve, and investigate how the maximum energy to which
particles are accelerated, and its time evolution, depends on various
parameters. We take into account the ambient magnetic field and its
amplification by resonant or non-resonant modes. Our studies reveal
that the maximum energy to which particles are accelerated is
generally reached in the ejecta-dominated stage, much before the start
of Sedov stage.  For SNe evolving within the winds of their massive
stars, the maximum energy is reached very early in the evolution. We
briefly explore the consequences for supernova remnants expanding in
surroundings such as wind-bubbles or superbubbles.

\end{abstract}


\end{frontmatter}


\section{Introduction}
\label{sec:intro}

Supernova remnants (SNRs) have long been thought to be responsible for
accelerated particles, at least up to the knee of the cosmic-ray
spectrum. This is usually thought to occur by some process related to
Diffusive Shock Acceleration \cite[DSA,][]{drury83, md01} and its
nonlinear modification \cite{edm97}. However many details of the
process are still being worked out.  Issues under active study include
the diffusion coefficient at the shocks and beyond the shock front,
the maximum energies to which particles can be accelerated and the
factors that determines this, and the radiative signatures from the
remnant. A complete understanding of the system requires that the
acceleration of particles at SNR shocks must be studied in conjunction
with the escape of the particles from the accelerator. In this paper
we discuss some details about the escape of particles from SNRs, and
especially how the dynamics and kinematics of young SNRs influence the
maximum energy to which particles are accelerated.

\section{Escape of Particles from Young Supernova Remnants}
\label{sec:escape}

An overview of the escape of particles from SNRs, and issues related
to this problem, has been recently discussed by Drury \cite{drury11},
which explores the many different ideas that exist. In many
discussions of young SNRs and particle escape, it is assumed that the
SNR velocity is constant in the so-called free-expansion stage, (which
we refer to here more appropriately as the ejecta-dominated stage),
and that no particles escape from the SNR in this stage
\cite{cba09}. The argument supposes that the maximum energy of
particles increases until the start of the Sedov or adiabatic stage,
and that particles start escaping only in the Sedov stage \cite{pz03,
  gabici11, helderetal12}. Caprioli et al.~\cite{cba09} insist that
decreasing maximum momentum or energy of the particles will result in
all particles which have energy above the maximum at that timestep
leaving the system, and therefore associate escape with decreasing
maximum momentum, which they claim happens only in the Sedov
phase. Gabici \cite{gabici11} asserts that a decelerating spherical
shock wave can result in particles escaping from the shock. If the
shock velocity is decreasing with time, the length over which the
particles are diffusing $ L_{diff} \propto D / v_{sh}$ ($D$ is the
diffusion coefficient, and $v_{sh}$ is the shock velocity), is
increasing with time, thus leading to eventual escape from the
accelerator.

Ellison \& Bykov \cite{eb11} take the view that escape of particles is
a fundamental part of the acceleration process that is going to take
place at any given stage regardless of maximum momentum or time
evolution. Our numerical calculations using test particles
\cite{tdp12b} tend to agree with \cite{eb11}. In this paper we support
the view that escape will occur even at an early stage. For most
reasonable models of SNR expansion, the shock velocity is always
decreasing. The maximum energy is decreasing in many of the more
common situations. Previous results that showed otherwise are flawed
because they make inaccurate assumptions about the SNR evolution.

The assumptions that are made in this paper regarding SNR expansion
differ in two main ways from many other calculations (1) The shock
velocity is taken from actual young SNR expansion models (rather than
an ad-hoc invocation that the velocity is approximately constant as
used in many previous analyses). As shown by Chevalier
\citep{chevalier1982a,chevalier1994}, the expansion of a young SNR
into a surrounding medium, however tenuous, results in a shock
velocity that is {\em always} decreasing with time. (2) The SNR takes
much longer to reach the Sedov stage than is generally assumed. A
general assumption that is made is that SNRs reach the Sedov stage
when the swept-up mass equals the mass ejected in the
explosion. However, since at least the work of Gull \cite{gull73}, it
has been known that this is not quite true, and the swept-up mass must
exceed the ejected mass significantly before the remnant can be
assumed to have the characteristics of the Sedov stage. This was
quantified further by Dwarkadas \& Chevalier \cite{dc98}, for
different ejecta density profiles, where it was shown that the mass of
swept-up material must exceed the ejected mass by a factor of 15-30
before the remnant can be considered to be in the Sedov stage. The
important point here is that it takes a much longer time to reach the
Sedov stage \cite{dwarkadas11a}, perhaps thousands of years in a low
density medium, and that while the remnant is in the ejecta-dominated
stage, its velocity will gradually decrease. Our goal here is to
study, via simple analytic expansions, how this SNR evolution in the
early stages affects the evolution of the maximum energy of
accelerated particles.

\section{Maximum Energy Considerations}
\label{sec:maxen}
The expansion of a SNR results in the formation of a shock wave that
expands with very high Mach number into the surrounding
medium. Particles are assumed to be accelerated at the collisionless
shock wave (see Spitkovsky, HEDLA 2012 proceedings) to relativistic
energies by some process, which is generally thought to be diffusive
shock acceleration (DSA) \cite{drury83}. In order to understand the
maximum energy that an accelerated particle can attain, we must first
recognize the processes that limit the energy that the accelerated
particle can attain.

The acceleration time for particles undergoing DSA can be written
approximately as

\be
t_{acc} \propto D / v_{sh}^2
\label{eq:tacc}
\ee

\noindent
where $D$ is the diffusion coefficient, and $v_{sh}$ is the shock
velocity (which could relate to either the forward or reverse
shock). Factors of order unity that do not change with time are left
out. In the vicinity of shocks, the scattering of particles via
magnetic irregularities is so efficient that the diffusion coefficient
can decrease to that corresponding to Bohm diffusion, where the
particle mean free path is of the order of the Larmor radius. This is
potentially the smallest value that the diffusion coefficient could
have. In this case equation~\ref{eq:tacc} can be written as:

\be
t_{acc} \propto E/ B\;v_{sh}^2
\label{eq:tacc}
\ee

\noindent
where $E$ is the energy of the particle, and $B$ the magnetic
field. One obvious limitation on the maximum energy then is that it
corresponds to the age of the remnant, since particles cannot be
accelerated for any greater time. Therefore

\be
E_{max} \propto B\;v_{sh}^2\;t_{age}
\label{eq:emax}
\ee 

\noindent
where $t_{age}$ is the age of the remnant. More accurate calculations
\cite{drury01} suggest that there is an additional factor of order
unity (\cite{drury01} gives 0.3), but as long as it is not time
dependent we can ignore it here. There may be other limitations, such
as a maximum wavelength of the scattering waves outside the shock
\cite{reynolds11}, or that the cosmic ray scale height cannot exceed
the SNR radius \cite{bl01}. Electrons may also be subject to loss
mechanisms such as synchrotron losses, that further reduce the maximum
energy. In this paper however we will concentrate on the major
limitation due to age, and further assume that the maximum energy of
escaped particles is the maximum energy to which particles are
accelerated, which is a reasonable assumption.

The evolution of a young supernova remnant (SNR) has been described in
many papers. We use the formulation suggested by Chevalier
\citep{chevalier1982a,chevalier1994}. In brief, the expansion of SN
ejecta into the surrounding medium leads to the formation of a
double-shocked structure, consisting of a reverse shock that travels
back into the ejecta, and a forward shock that expands into the
ambient medium. If the SN ejecta are described by ${\rho}_{ej} \propto
r^{-n}$, and the surrounding medium is described by ${\rho}_{amb}
\propto r^{-s}$, then the self-similar solutions
\citep{chevalier1982a} show that the contact discontinuity will expand
as:

\be
R_{CD} \propto t^{(n-3)/(n-s)}
\ee

Since the expansion is self-similar, the forward and reverse shocks
will expand in the same manner. We can write the expansion of the
forward shock as R$_f \propto t^m$, where m=(n-3)/(n-s) is referred to
as the expansion parameter. Note that since the solutions require $n >
5$, and $s < 3$, we have $m \leq 1$. The velocity $v_{sh} \propto
t^{m-1}$, and is always decreasing with time. This suggests, according
to one of the arguments above, that particles must be escaping the
system.

Going back to eqn.~\ref{eq:emax}, we can then write that

\be
E_{max} \propto B\;t^{2m-2}\;t_{age} \propto B\;t^{2m-1}
\label{eq:emaxv}
\ee 

It is clear that, as long as $m > 0.5$, and the magnetic field $B$ is
constant, the maximum energy is increasing with time. The maximum
energy will start to decrease with time once $m < 0.5$, i.e. just
before the remnant enters the Sedov stage. This is somewhat consistent
with what other authors have postulated in the past
\cite{pz05,gabici11}, although it indicates that the maximum energy
will be reached sometime before the remnant enters the Sedov
stage. Furthermore, since the value of $m$ is constant in the
self-similar case, the energy will increase at the same rate for much
of the self-similar phase, before $m$ starts to decrease as the
remnant enters the Sedov stage.

The crucial point here is with regards to the behaviour of the
magnetic field. There are several indications that the field measured
from radio, X-ray and gamma-ray observations of SNRs far exceeds the
field in the general interstellar medium. The general topic of
magnetic fields in cosmic particle acceleration sources is
comprehensively reviewed in \cite{ber12}. Here we concentrate on two
main issues (1) What is the value of the ambient magnetic field and
(2) How does the amplified magnetic field behave?

\subsection{Ambient magnetic field}

If the SN shock is expanding in the general interstellar medium, then
one can assume, at least as a first approximation, that the magnetic
field is equal to the interstellar magnetic field, with a value around
5 $\mu$G. This may be the case for Type Ia SNe, whose progenitors are
supposed to be white dwarf stars that do not significantly modify the
medium around them.

However, all other SNe arise from the core-collapse of massive
stars. These massive stars suffer serious mass-loss throughout their
lifetimes, losing a large fraction of their initial mass (see
Dwarkadas, HEDLA 2012 proceedings). After the SN explodes, the shock
wave will evolve in the medium crafted out by the wind from the star,
and not in the interstellar medium.

In order of increasing radius from the star, the wind blown medium
surrounding the star consists of \cite{Weaver1977, Dwarkadas2005,
  Dwarkadas2007c} (1) A freely expanding wind, whose density decreases
as r$^{-2}$ if the wind parameters are constant (2) A wind termination
shock (3) A low density, hot shocked wind medium (4) A contact
discontinuity (5) Shocked ambient medium (6) Wind shock (7) unshocked
ambient medium, be it another wind or the interstellar medium. This
picture can easily be distorted by winds whose parameters change with
time, turbulence, instabilities etc. But in general the SN shock
should first be evolving in the wind of the progenitor star.

For a star that is rotating, the field lines near a star resemble an
Archimedean spiral, with the radial field falling as r$^{-2}$ and the
tangential as r$^{-1}$. Thus the tangential component will dominate at
large radii, and the field in the wind can be assumed to decrease
inversely with radius. If $B \propto r^{-1}$ then

\be
E_{max} \propto  r^{-1}\;t^{2m-1} \propto t^{-m}\;t^{2m-1} \propto t^{m-1}
\label{eq:emaxwind}
\ee 

Since $m < 1$, in the case of a SNR evolving in a wind E$_{max}$ is
always decreasing with time! Since $\sim$80\% of SNRs arise from
massive star progenitors, we expect that this will be the predominant
case. For a SN with ejecta density decreasing as r$^{-9}$ expanding
into a wind, m $\sim$ 0.86, and the maximum energy will decrease as
t$^{-0.14}$. Thus the maximum energy is a slowly decreasing function
of time.

\subsection{Amplified Magnetic field}

\subsubsection{Non-resonant modes} 

In 2001, Bell \& Lucek \cite{bl01} showed that the magnetic field can
be amplified non-linearly by the cosmic rays themselves, to
significantly exceed the pre-shock value. Early in the evolution,
non-resonant modes dominate, while later on resonant modes seem to be
more dominant \cite{cba09}. If the non-resonant modes dominate, the
maximum magnetic field is obtained when the magnetic energy density is
equal to the kinetic energy density. In this case the resonantly
amplified magnetic field is given by \cite{cba09}

\be
B_{amp, nr} = \sqrt{2 \pi \rho_{amb} \;(v_{sh}^3 /c) \xi}
\ee

\noindent
where $\xi$ is an acceleration efficiency. Assuming $\xi$ to be a
constant (not necessarily the case), we get that $B_{amp} \propto
\rho_{amb}^{0.5}\;t^{3(m-1)/2}$. Therefore, in a constant density
medium we will have $B_{amp} \propto t^{3(m-1)/2}$. In a wind medium
with constant mass-loss parameters, $\rho_{amb} \propto r^{-2}$, and
we have $B_{amp} \propto r^{-1}\;t^{3(m-1)/2} \propto t^{(m-3)/2} $.

This gives, for a SNR evolving in a constant density medium with a
non-resonantly amplified magnetic field, that

\be
E_{max, NR} \propto t^{3(m-1)/2}\;t^{2m-1} \propto t^{(7m-5)/2}
\label{eq:emaxamp}
\ee 

Thus for $m < 5/7 = 0.71$ the energy decreases with time, while for
larger $m$ it increases with time. For n=9, we get that $m=0.66$ for a
constant density medium (s=0), and the maximum energy will decrease
with time, whereas for n=11 it will be very slowly increasing.

If the SNR is evolving in a wind medium we obtain:

\be
E_{max, NR, w} \propto t^{(m-3)/2}\;t^{2m-1} \propto t^{(5m-5)/2}
\label{eq:emaxampw}
\ee 

In this case, irrespective of the value of m, the maximum energy is
always decreasing with time.

\subsubsection{Resonant Modes}

If resonant modes dominate, as is more likely later in the evolution, then

\be
B_{amp, r} = \sqrt{8 \pi \rho_{amb} \;(v_{sh}^2) \xi/M_A}
\ee

\noindent
where $M_A$ is the Mach number. Given that $M_A = v_{sh}/v_A$ where
$v_A = B_{amb}/(\sqrt{4 \pi \rho})$ is the Alfven velocity, we have:

\be
B_{amp, r} \propto \sqrt{B_{amb}\;{\rho_{amb}}^{1/2} \;(v_{sh})\; \xi}
\ee

Thus we get that $B_{amp,r} \propto
{B_{amb}}^{0.5}\;\rho_{amb}^{0.25}\;t^{(m-1)/2}$. Therefore, in a
constant density medium we will have $B_{amp,r} \propto
t^{(m-1)/2}$. In a wind medium with constant mass-loss parameters,
$\rho_{amb} \propto r^{-2}$, $B_{amb} \propto r^{-1}$ and we have
$B_{amp,r} \propto r^{-1/2}\;r^{-1/2}\;t^{(m-1)/2} \propto t^{-(m+1)/2}
$.

Again, making the assumption that $\xi$ is constant with time, we get
for the maximum energy in a constant density medium that:

\be
E_{max, R} \propto t^{(m-1)/2}\;t^{2m-1} \propto t^{(5m-3)/2}
\label{eq:emaxampr}
\ee 

Thus for $m < 3/5 = 0.6$ the energy decreases with time, while for
larger $m$ it increases with time. For n=9, we get that $m=0.66$ and
the maximum energy will increase with time, whereas for n=7 it will
decrease with time.

For a wind medium we get that

\be
E_{max, R} \propto t^{-(1+m)/2}\;t^{2m-1} \propto t^{3(m-1)/2}
\label{eq:emaxamprw}
\ee 

\noindent 
which is always decreasing with time.

\section{Discussion}

In Table 1 below we summarize the dependence of the evolution of
maximum energy on various assumptions of the magnetic field in young
supernova remnants.

\begin{center}
\begin{tabular}{|l|c|c|c|c|} \hline
 & Unamplified Field & Amplified Field (Non-resonant)  & Amplified Field (Resonant) \\ \hline
Constant density medium & t$^{2m-1}$ & t$^{(7m-5)/2}$ & t$^{(5m-3)/2}$   \\ \hline
Wind Medium $\rho \propto r^{-2}$ & t$^{m-1}$  & t$^{5(m-1)/2}$ & t$^{3(m-1)/2}$ \\ \hline

\end{tabular}
\end{center}

As is clear from this table, in the case of a wind-medium, the maximum
energy {\em in all cases} is always decreasing with time. In the case
of a constant density medium, the energy may or may not decrease in
the initial stages depending on the value of the expansion
parameter. However, as the SNR approaches the Sedov-Taylor phase, the
value of $m$ will drop and the energy will begin to decrease. Note
that in every case, the energy begins to decrease at a value of $m >
0.4$, i.e. before it reaches the Sedov value. The velocity is {\em
  always} decreasing with time in this ejecta-dominated stage. This
will be accompanied by an escape of particles from the SNR even in
this early stage, as confirmed by our more precise numerical
calculations \cite{tdp12a, tdp12b}.

We have considered a few likely possibilities for the magnetic field,
and used it to derive the above results. Other magnetic field
behaviour is quite possible, and may lead to a different evolution of
the magnetic field and therefore the maximum energy. What is clear is
that the {\em big unknown in the evolution of the maximum energy is
  the magnetic field}. Although surprises still abound, we know to a
much better precision how SNRs expand in the ambient medium (and thus
the values of $R$ and $v_{sh}$ in equation \ref{eq:emax}), than we
know how the magnetic field behaves (and thus the value of $B$ in
equation \ref{eq:emax}). The value of the maximum energy, and whether
SNRs can accelerate particles to the ``knee'' of the cosmic ray
spectrum and beyond, also depends crucially on the magnetic field (see
also \cite{drury11}).

Although this calculation uses simple arguments that do not take into
consideration any of the intricacies of the acceleration process, it
gives a general idea of the evolution of maximum energy. We have also
assumed for simplicity that the acceleration efficiency is constant,
which is possibly not the case. Our main point though is to emphasize
the evolution of maximum energy of accelerated particles in the ejecta
dominated stage, and demonstrate that it leads to a conclusion that
contradicts much of the previous work (see Fig 2 in
\cite{helderetal12}) - the maximum energy of SNRs must be reached
somewhere in the ejecta dominated phase, in many cases much before the
SNR reaches the Sedov or adiabatic phase, and not at the beginning of
the Sedov phase. This agrees with numerical calculations \cite{eb11,
  tdp12b}.

In the case of a wind medium, the results indicate that the maximum
energy is always decreasing. This obviously cannot be extrapolated
back to time $t=0$ as the time of maximum energy, because one must
take into account the finite time for particles to carry out several
crossing of the shock front and reach maximum energy. The energy will
continue to increase during this time, then reach a maximum after
which the above results come into play. This time of increase though
could be short, maybe a few tens to hundred of years (see Fig
\ref{fig:emax}).

These results lead to some surprising inferences. Core-collapse SNRs
may begin their evolution within a wind medium, but do not continue to
evolve in a wind medium for a long time. The SN will first evolve in
the wind of the star, and subsequently, after crossing the
wind-termination shock, in an almost constant density medium. What the
afore-mentioned results then suggest is that when the SNR is evolving
in the stellar wind, the maximum energy of the particles must be
decreasing slowly, but when it starts evolving in the constant density
medium the maximum energy must begin to increase with time. This would
mean that in wind-blown bubbles, and in their larger cousins,
superbubbles, the energy would first decrease and then increase,
before decreasing again as the SNR evolves to the Sedov stage. The
maximum of the energy could be reached (depending on the actual
parameters) later in the evolution while the SNR is evolving in a hot,
low density shocked wind.  The higher energy particles arise in the
shocked wind, whereas the lower energy ones could arise from the
unshocked wind, with possibly different compositions.

\section{Comparison with Observations:} 

As outlined above, many approximations were made in obtaining these
results. Also, these are technical upper limits, probably applicable
only to protons as electrons will experience radiative losses that
reduce the energy.  In some cases they may be superceded by other
limits.  In order to investigate the evolution of maximum energy in
detail, we have calculated it in one such situation via detailed
numerical calculations that take the acceleration process and energy
losses into account. Our methodology is described in two papers by our
group \cite{tdp12a, tdp12b}. The specific calculation below however is
unpublished and will be described in a subsequent paper. In this
particular case, a Type Ic SNR was simulated expanding first in a wind
and then, after crossing the wind termination shock, in a constant
density medium. No magnetic field amplification was assumed - the
field falls off as r$^{-1}$ in the wind, and is constant in the
shocked region. Note that interaction of the SN shock with the wind
termination shock will result in a compression of the SN shocked
region and an appropriate increase in the magnetic field.

\begin{figure}[ht]
\includegraphics[scale=0.9]{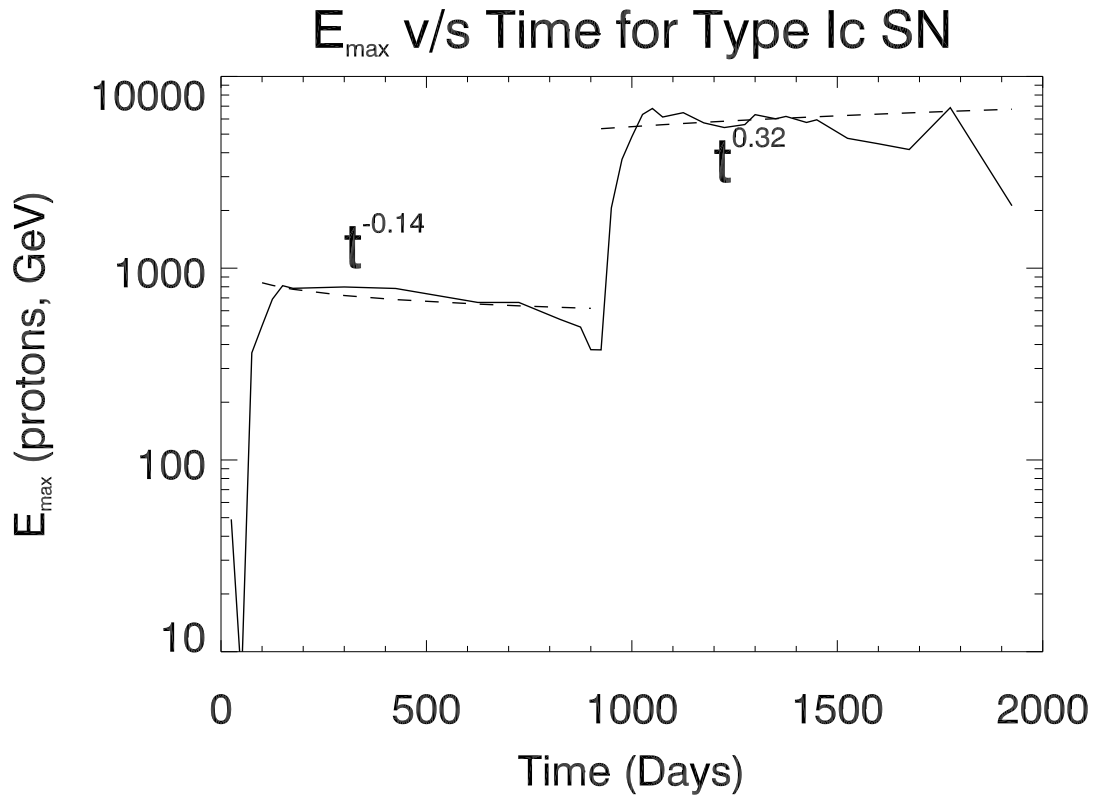}
\caption{The maximum energy versus time for a Type Ic SNR evolving in
  a stellar wind, followed by a hot shocked medium. The solid line is
  the energy computed from a direct calculation involving acceleration
  at both shocks, the dashed line is the values given by our analytic
  approximations. Our theory predicts that the maximum energy should
  decrease during the evolution in the wind, and it seems to agree
  quite well with that observed in the numerical simulations. The
  later evolution is not self-similar and therefore the theoretical
  model does not work as well. \label{fig:emax}}
\end{figure}

Figure \ref{fig:emax} shows the maximum energy of protons calculated
from the simulations (solid lines) and compared to the evolution
predicted from our calculations. The maximum energy of accelerated
particles is not an easy quantity to extract. We have computed the
maximum energy at time intervals of 25 years, plotted the points and
smoothed the curve to reflect the overall evolution. In the early
evolution, while the SNR, with $n=9$ is evolving in a wind, our
scalings predict that the maximum energy should decrease as
t$^{-0.14}$. We find that, after a finite period as described above
when the particles reach maximum energy, the energy does decrease, and
the slope does not differ substantially from the analytic value. This
speaks for the validity of our assumptions.

After crossing the wind shock, the energy is predicted to increase as
t$^{0.32}$, but seems instead to fluctuate. This is not difficult to
understand. The predictions are made for a purely self-similar
evolution. However, after crossing the wind shock, the SNR shocked
interaction region is disturbed and is not evolving in a self-similar
fashion, although it will approach self-similarity after a few
crossing times. In a sense the shock profile is intermediate between
that of a SNR in a wind and in a constant density medium, although
eventually it will approach that of a SNR in a constant density
medium, unless the nature of the medium changes before it does
so. Although our simulation has not been carried on for that long,
when the evolution becomes self-similar again we should see the
increase that is predicted. 

These results are not so easily applicable to Type Ia SNRs, because
their ejecta profiles are more complicated. \cite{dc98} showed that
the ejecta profile could be approximately fit by an exponential
density. Unfortunately, such a profile is not self-similar, and does
not lend itself easily to analytic calculations.

\section{Summary and Conclusions} 

As mentioned earlier, many authors have neglected the variation of the
SNR in the early phase, assuming that the velocity is constant in the
so-called free-expansion phase, that the maximum energy of accelerated
particles is always increasing in this phase, and that no particles
escape. In this paper we have started with the velocity profile from
self-similar solutions of SNR evolution, which show that the velocity
in the ejecta-dominated phase is always decreasing with time. We have
used that to compute the evolution of the maximum energy of the
escaping particles assuming Bohm diffusion at the SNR shock. We have
shown that the maximum energy is always decreasing (after an initial
short period) in the case of SNRs evolving in winds, whereas it may
increase, but eventually begin to decrease before the start of the
Sedov stage, in case the SNR is evolving in a constant density
medium. The big unknown in each case is the magnetic field
evolution. These simple results lead to some surprising implications
for the evolution of the maximum energy of accelerated particles at
SNR shocks in various complicated surroundings. In a follow-up paper
we will consider other limiting cases, other possibilities for the
magnetic field, and quantify the various approximations to show how
the maximum energy depends on various SNR parameters. We will also
quantify what the maximum energy is for various SN types. In future,
we will study these more accurately using numerical simulations.












\end{document}